\documentstyle[11pt]{article}
\begin{document}

\title{{\bf The generalized second law in irreversible thermodynamics for the interacting dark energy
in a non-flat FRW universe enclosed by the apparent horizon}}
\author{K. Karami$^{1,2}$,\thanks{E-mail: KKarami@uok.ac.ir}\\
S. Ghaffari${^1}$\\$^{1}$\small{Department of Physics, University of
Kurdistan, Pasdaran St., Sanandaj, Iran}\\$^{2}$\small{Research
Institute for Astronomy $\&$ Astrophysics of Maragha (RIAAM),
Maragha, Iran}\\}

\maketitle

\begin{abstract}
We investigate the validity of the generalized second law in
irreversible thermodynamics in a non-flat FRW universe containing
the interacting dark energy with cold dark matter. The boundary of
the universe is assumed to be enclosed by the dynamical apparent
horizon. We show that for the present time, the generalized second
law in nonequilibrium thermodynamics is satisfied for the special
range of the energy transfer constants.
\end{abstract}
\clearpage
\section{Introduction}
Type Ia supernovae observational data suggest that the universe is
dominated by two dark components: dark matter and dark energy
\cite{Riess}. Dark matter (DM), a matter without pressure, is mainly
used to explain galactic curves and large-scale structure formation,
while dark energy (DE), an exotic energy with negative pressure, is
used to explain the present cosmic accelerating expansion. However,
the nature of DE is still unknown, and people
 have proposed some candidates to describe it. The cosmological
 constant, $\Lambda$, is the most obvious theoretical candidate of
 DE, which has the equation of state $\omega=-1$.
 Astronomical observations indicate that the cosmological constant is many orders of magnitude
 smaller than estimated in modern theories of elementary particles \cite{Weinberg}. Also the
 "fine-tuning" and the "cosmic coincidence" problems are the two
 well-known difficulties of the cosmological constant problems
 \cite{Copeland}.

 There are different alternative theories for the dynamical DE scenario which have been
 proposed by people to interpret the accelerating universe. i) The scalar-field models of DE including
 quintessence \cite{Wetterich},
 phantom (ghost) field \cite{Caldwell1}, K-essence \cite{Chiba}
 based on earlier work of K-inflation \cite{Picon3},
 tachyon field \cite{Sen}, dilatonic ghost condensate \cite{Gasperini},
 quintom \cite{Elizalde}, and so forth. ii) The DE models including
 Chaplygin gas \cite{Kamenshchik}, braneworld models
 \cite{Deffayet}, holographic DE models
 \cite{Cohen}, and agegraphic DE models \cite{Cai}, etc.

Besides, as usually believed, an early inflation era leads to a
flat universe. This is not a necessary consequence if the number
of e-foldings is not very large \cite{Huang}. It is still possible
that there is a contribution to the Friedmann equation from the
spatial curvature when studying the late universe, though much
smaller than other energy components according to observations.
Therefore, it is not just of academic interest to study a universe
with a spatial curvature marginally allowed by the inflation model
as well as observations. Some experimental data have implied that
our universe is not a perfectly flat universe and that it
possesses a small positive curvature \cite{Bennett}.

In the semiclassical quantum description of black hole physics, it
was found that black holes emit Hawking radiation with a temperature
proportional to their surface gravity at the event horizon and they
have an entropy which is one quarter of the area of the event
horizon in Planck unit \cite{Hawking}. The temperature, entropy and
mass of black holes satisfy the first law of thermodynamics
\cite{Bardeen}. On the other hand, it was shown that the Einstein
equation can be derived from the first law of thermodynamics by
assuming the proportionality of entropy and the horizon area
\cite{Jacobson}. The study on the relation between the Einstein
equation and the first law of thermodynamics has been generalized to
the cosmological context where it was shown that the first law of
thermodynamics on the apparent horizon $R_{h}$ can be derived from
the Friedmann equation and vice versa if we take the Hawking
temperature $T_{h} =1/2\pi R_{h}$ and the entropy $S_{h} = \pi
R_{h}^2$ on the apparent horizon \cite{Cai05}. Furthermore, the
equivalence between the first law of thermodynamics and Friedmann
equation was also found for gravity with Gauss-Bonnet term and the
Lovelock gravity theory \cite{Cai05,Akbar}.

Besides the first law of thermodynamics, a lot of attention has
been paid to the generalized second law (GSL) of thermodynamics in
the accelerating universe driven by dark energy. The generalized
second law of thermodynamics is as important as the first law,
governing the development of the nature
\cite{Davies,Gong,Izquierdo2,Izquierdo,Zhau,Gong5,Wang2,Sheykhi3}.
For the different DE models like the generalized Chaplygin gas
\cite{Gong5}, the holographic DE \cite{Wang2,Sheykhi4},
 the braneworld scenarios \cite{Sheykhi2}, and the
new agegraphic DE \cite{KA}, people showed that the GSL for the
universe containing the DE and DM enclosed by the dynamical
apparent horizon is always satisfied throughout the history of the
universe for any spatial curvature.

Note that in \cite{Wang2}, the authors investigated the validity
of the first and the generalized second law of thermodynamics for
both apparent and event horizon for the case of holographic DE
with DM in a flat universe. They showed that in contrast to the
case of the apparent horizon, both the first and second law of
thermodynamics break down if one consider the universe to be
enveloped by the event horizon with the usual definitions of
entropy and temperature. They argued that the break down of the
first law can be attributed to the possibility that the first law
may only apply to variations between nearby states of local
thermodynamic equilibrium, while the event horizon reflects the
global spacetime properties. Author of \cite {Karami} cleared that
even by redefining the event horizon measured on the sphere of the
horizon as the system's IR cut-off for a holographic DE model in a
non-flat universe, the GSL cannot be satisfied.

One of another interesting issue in the cosmological context, is
the study of interaction between DE and DM. The choice of the
interaction between both components was to get a scaling solution
to the coincidence problem such that the universe approaches a
stationary stage in which the ratio of DE and DM becomes a
constant \cite{Hu}. Das et al. \cite{Das} computed leading-order
corrections to the entropy of any thermodynamic system due to
small statistical fluctuations around equilibrium. They obtained a
general logarithmic correction to black hole entropy. Wang et al.
\cite{Wang4} studied a thermodynamical description of the
interaction between holographic DE and DM. Resorting to the
logarithmic correction to the equilibrium entropy \cite{Das} they
 arrived to an expression for the interaction term which was
consistent with the observational tests. Pav\'{o}n and Wang
\cite{Pavon} considered a system composed of two subsystems (DM
and DE) at different temperatures. In virtue of the extensive
property, the entropy of the whole system is the sum of the
entropies of the individual subsystems which (being equilibrium
entropies) are just functions of the energies of DE and DM even
during the energy transfer process. Zhou et al. \cite{Zhou1} have
further employed the second law of thermodynamics to study the
coupling between the DE and DM in the universe by resorting to the
nonequilibrium entropy of extended irreversible thermodynamics.

Here our aim is to extend the work of Zhou et al. \cite{Zhou1} and
investigate the validity of the generalized second law in
irreversible thermodynamics for the interacting DE with DM in a
non-flat FRW universe enclosed by the dynamical apparent horizon.
The apparent horizon is more appropriate than the event horizon on
defining the thermodynamics in the universe \cite{Gong5}. This
paper is organized as follows. In Section 2, we study the DE model
in a non-flat universe which is in interaction with the cold DM.
In Section 3, we investigate the validity of the generalized
second law in irreversible thermodynamics for the universe
enclosed by the apparent horizon. Section 4 is devoted to
conclusions.
\section{Interacting DE and DM}

We consider the Friedmann-Robertson-Walker (FRW) metric for the
non-flat universe as
\begin{equation}
{\rm d}s^2=-{\rm d}t^2+a^2(t)\left(\frac{{\rm
d}r^2}{1-kr^2}+r^2{\rm d}\Omega^2\right),\label{metric}
\end{equation}
where $k=0,1,-1$ represent a flat, closed and open FRW universe,
respectively. Observational evidences support the existence of a
closed universe with a small positive curvature ($\Omega_{k}\sim
0.02$) \cite{Bennett}. The first Friedmann equation for the
non-flat FRW universe takes the form
\begin{equation}
{\textsl{H}}^2+\frac{k}{a^2}=\frac{8\pi}{3}~ (\rho_{x}+\rho_{\rm
m}),\label{eqfr}
\end{equation}
where we take $G=1$. Also $\rho_{x}$ and $\rho_{\rm m}$ are the
energy density of DE and DM, respectively. Let us define, as usual,
the fractional energy densities as
\begin{equation}
\Omega_{\rm m}=\frac{\rho_{\rm m}}{\rho_{\rm
cr}}=\frac{8\pi\rho_{\rm m}}{3H^2},~~~~~~\Omega_{\rm
x}=\frac{\rho_{x}}{\rho_{\rm
cr}}=\frac{8\pi\rho_{x}}{3H^2},~~~~~~\Omega_{k}=\frac{k}{a^2H^2},
\label{eqomega}
\end{equation}
then, the first Friedmann equation can be written as
\begin{equation}
\Omega_{\rm m}+\Omega_{x}=1+\Omega_{k}.\label{eq10}
\end{equation}
We consider a universe containing an interacting DE density
$\rho_{x}$ and the cold DM (CDM), with $\omega_{\rm m}=0$. The
energy equations for DE and CDM are
\begin{equation}
\dot{\rho}_{x}+3H(1+\omega_{x})\rho_{x}=-\Gamma,\label{eqpol}
\end{equation}
\begin{equation}
\dot{\rho}_{\rm m}+3H\rho_{\rm m}=\Gamma,\label{eqCDM}
\end{equation}
where following \cite{Zhou1}, we choose $\Gamma=3H\lambda
\rho_{x}$ as an interaction term with $\lambda$ a small,
dimensionless, positive quantity. For $\Gamma>0$ the energy
proceeds from DE to DM. Although this expression for the
interaction term may look purely phenomenological but different
Lagrangians have been proposed in support of it \cite{Tsujikawa}.
Note that choosing the $H$ in the $\Gamma$-term is motivated
purely by mathematical simplicity. Because from the continuity
equations, the interaction term should be proportional to a
quantity with units of inverse of time. For the latter the obvious
choice is the Hubble factor $H$. The dynamics of interacting DE
models with different $\Gamma$-classes have been studied in ample
detail by \cite{Amendola}. It should be emphasized that this
phenomenological description has proven viable when contrasted
with observations, i.e., SNIa, CMB, large-scale structure, $H(z)$,
and age constraints \cite{Wang8}, and recently in galaxy clusters
\cite{Bertolami8}.

The deceleration parameter is given by
\begin{equation}
q=-\Big(1+\frac{\dot{H}}{H^2}\Big).\label{q1}
\end{equation}
Taking the time derivative in both sides of Eq. (\ref{eqfr}), and
using Eqs. (\ref{eqomega}), (\ref{eq10}), (\ref{eqpol}) and
(\ref{eqCDM}), we get
\begin{equation}
q=\frac{1}{2}\Big(1+\Omega_{k}+3\Omega_{x}\omega_{
x}\Big).\label{q2}
\end{equation}

\section{Generalized second law in irreversible thermodynamics}
Here, we study the validity of the generalized second law (GSL) in
irreversible (nonequilibrium) thermodynamics. According to the GSL
in equilibrium thermodynamics, entropy of matter and fluids inside
the horizon plus the entropy of the horizon do not decrease with
time \cite{Wang2}. In equilibrium thermodynamics, irreversible
fluxes such as energy transfers play no part and they do not enter
the entropy function which is defined for equilibrium states only.
However, in nonequilibrium extended thermodynamics such fluxes
enter the entropy function \cite{Zhou1}.

The entropy of the DE and CDM which are in interaction with each
other are given by Gibb's equation \cite{Izquierdo,Zhau,Wang2}
\begin{equation}
T_x{\rm d}S_x={\rm d}Q_x={\rm d}E_x+P_x{\rm d}V,\label{eqSLT1}
\end{equation}
\begin{equation}
T_m{\rm d}S_m={\rm d}Q_m={\rm d}E_m,\label{eqSLT2}
\end{equation}
where like \cite{Wang2}, ${\rm V}=4\pi R_{\rm h}^3/3$ is the
volume containing the DE and CDM with the radius of the horizon
$R_{\rm h}$. And
\begin{equation}
E_x=\rho_{x}V,~~~E_m=\rho_{\rm m}V,\label{eqEde}
\end{equation}
\begin{equation}
P_{x}=\omega_{x}\rho_{x}=\frac{3H^2}{8\pi}\omega_{x}\Omega_{x}.
\label{eqEcdm}
\end{equation}
Also $T_x$ and $T_m$ are the temperatures of DE and CDM,
respectively, and following \cite{Zhou1} are given by
\begin{equation}
T_{x}=T_{eq}e^{-3(\omega_{x}+\lambda)(x-x_{eq})},\label{eqTx}
\end{equation}
\begin{equation}
T_{m}=T_{eq}\frac{r}{r_{eq}}e^{-[2+3(\omega_{x}+\lambda)](x-x_{eq})},\label{eqTm}
\end{equation}
where $r=\Omega_{m}/\Omega_{x}$ and $x=\ln a$. The subscript '$ eq$'
indicates the value taken by the corresponding quantity when DE and
DM are in thermal equilibrium. Note that in Eqs. (\ref{eqTx}) and
(\ref{eqTm}), $\lambda$ and $\omega_x$ are constants. We limit
ourselves to the assumption that the thermal system including the DE
and CDM bounded by the horizon in the absence of interaction remain
in equilibrium so that the temperature of the system must be uniform
and the same as the temperature of its boundary. This requires that
the temperature $T$ of the both DE and CDM inside the horizon in the
absence of interaction should be in equilibrium with the Hawking
temperature $T_{\rm h}$ associated with the horizon, so we have
$T_{eq}=T_{h}=1/(2\pi R_{h})$. In the presence of interaction, when
the temperature of the system differs from that of the horizon,
there will be spontaneous heat flow between
  the horizon and the fluid and the thermal equilibrium will no
longer hold \cite{Izquierdo,Zhau}.

Taking the derivative in both sides of (\ref{eqSLT1}) and
(\ref{eqSLT2}) with respect to cosmic time $t$, and using Eqs.
(\ref{eqfr}), (\ref{eqomega}), (\ref{eq10}), (\ref{eqpol}),
(\ref{eqCDM}), (\ref{eqEde}) and (\ref{eqEcdm}), we obtain
\begin{equation}
\dot{Q}_{x}=4\pi
R_{h}^2(\dot{R}_{h}-HR_{h})(1+\omega_{x})\rho_{x}-4\pi
R_{h}^3H\lambda \rho_{x},\label{Qx}
\end{equation}
\begin{equation}
\dot{Q}_{m}=4\pi R_{h}^2(\dot{R}_{h}-HR_{h})\rho_{m}+4\pi
R_{h}^3H\lambda \rho_{x}\label{Qm}.
\end{equation}
Also in addition to the entropies of DE and CDM in the universe,
there is a geometric entropy on the horizon $S_{h}=\pi R_{h}^{2}$
\cite{Wang2}. The evolution of this horizon entropy is obtained as
\begin{equation}
\dot{S}_h=2\pi R_h\dot{R_h}.\label{Sh}
\end{equation}
In the presence of interaction between DE and CDM inside the
universe enveloped by the horizon, the GSL in irreversible
 thermodynamics can be obtained by extending Eq.
(13) in \cite{Zhou1} as
\begin{equation}
\dot{S}^*=\frac{\dot{Q}_{m}}{T_{m}}+\frac{\dot{Q}_{x}}{T_{x}}
-A_x\dot{Q}_{x}\ddot{Q}_{x}-A_h\dot{Q}_{h}\ddot{Q}_{h}+\dot{S}_{h},\label{eqnonStot}
\end{equation}
where $A_x$ and $A_h$ are the energy transfer constants between DE
and DM inside the universe and between the universe and the
horizon, respectively.  Since the overall system containing the
universe and the horizon is isolated, one has
$\dot{Q}_{h}=-(\dot{Q}_{m}+\dot{Q}_{x})$. For an isolated universe
only, i.e. $\dot{Q}_{h}=0$ and $\dot{S}_{h}=0$, then Eq.
(\ref{eqnonStot}) reduces to Eq. (13) in \cite{Zhou1}.

Here we assume the boundary of the universe to be enveloped by the
dynamical apparent horizon. Hence, define $\tilde{r} = ar$, the
metric (\ref{metric}) can be rewritten as ${\rm d}s^2 = h_{ab}{\rm
d}x^a{\rm d}x^b + \tilde{r}^2{\rm d}\Omega^2$, where $x^a = (t,
r)$, $h_{ab}$ = diag($-1, a^2/(1 - kr^2)$). By definition,
$h^{ab}\partial_{a}\tilde{r}\partial_{b}\tilde{r} = 0$, the
location of the apparent horizon in the FRW universe is obtained
as $\tilde{r} = R_{h} = (H^2+k/a^2)^{-1/2}$ \cite{Cai09}. For $k =
0$, the apparent horizon is same as the Hubble horizon.

Cai \& Kim \cite{Cai05} proofed that the Friedmann equations in
Einstein gravity are derived by applying the first law of
thermodynamics to the dynamical apparent horizon, $R_{h}$, of a
FRW universe with any spatial curvature in arbitrary dimensions
and assuming that the apparent horizon has an associated entropy
$S_{h}$ and Hawking temperature $T_{h}$ as $S_{h} = \pi R_{h}^2$,
$T_{h} =1/2\pi R_{h}$. In the braneworld scenarios, the Friedmann
equations also can be written directly in the form of the first
law of thermodynamics, at the apparent horizon with the Hawking
temperature on the brane, regardless of whether there is the
intrinsic curvature term on the brane or a Gauss- Bonnet term in
the bulk \cite{Sheykhi2}. Recently the Hawking radiation with
temperature $T_{h} = 1/2\pi R_{h}$ on the apparent horizon of a
FRW universe with any spatial curvature has been observed in
\cite{Cai09}. The Hawking temperature is measured by an observer
with the Kodoma vector inside the apparent horizon \cite{Cai09}.

For the dynamical apparent horizon
\begin{equation}
R_{h}=H^{-1}(1+\Omega_{k})^{-1/2},\label{ah}
 \end{equation}
if we take the derivative in both sides of (\ref{ah}) with respect
to cosmic time $t$, then we obtain
\begin{equation}
\dot{R_{h}}=\frac{3(1+\Omega_{k}+\Omega_{x}\omega_{x})}{2(1+\Omega_{k})^{3/2}}\label{ahdot}.
\end{equation}
Using Eqs. (\ref{ah}) and (\ref{ahdot}) one can get
\begin{equation}
\dot{R_{h}}-HR_{h}=\frac{(1+\Omega_k+3\Omega_{x}\omega_{x})}{2(1+\Omega_{k})^{3/2}}\label{ahahdot}.
\end{equation}
Substituting Eqs. (\ref{ah}), (\ref{ahdot}) and (\ref{ahahdot}) in
Eqs. (\ref{Qx}), (\ref{Qm}) and (\ref{Sh}) reduce to
\begin{equation}
\dot{Q}_{x}=\frac{3}{2(1+\Omega_{k})^{5/2}}
\Big(q(1+\omega_{x})-\lambda(1+\Omega_{k})\Big)\Omega_{x},\label{eqdQx}
\end{equation}
\begin{equation}
\dot{Q}_{m}=\frac{3}{2(1+\Omega_{k})^{5/2}}
\Big(q\Omega_{m}+\lambda\Omega_{x}(1+\Omega_{k})\Big),\label{eqdQm}
\end{equation}
\begin{eqnarray}
\dot{S}_{h}=\frac{3\pi}{H(1+\Omega_{k})^2}(1+\Omega_{k}+\Omega_{x}\omega_{x}).\label{SAah}
\end{eqnarray}
Taking the time derivative of Eqs. (\ref{eqdQx}), (\ref{eqdQm}),
and using
\begin{equation}
\dot{\Omega_{k}}=2Hq\Omega_{k},
\end{equation}
\begin{equation}
\dot{\Omega_{x}}=H\Big[2(q+1)-3(1+\omega_{x}+\lambda)\Big]\Omega_{x},
\end{equation}
we get
\begin{eqnarray}
\ddot{Q}_{x}=\frac{3H}{2(1+\Omega_{k})^{5/2}}\Big\{\Big[-2(q+1)
+\frac{3q\Omega_{k}}{(1+\Omega_{k})}+3(1+\omega_{x}+\lambda)\Big]
\lambda(1+\Omega_{k})\nonumber\\+\Big[2q(q+1)-\frac{5q^2\Omega_{k}}{(1+\Omega_{k})}+\frac{\dot{q}}{H}-3q(1+\omega_{x}+\lambda)\Big](1+\omega_{x})
\Big\}\Omega_{x},\label{eqddQx}
\end{eqnarray}
\begin{eqnarray}
\ddot{Q}_{m}=\frac{3H}{2(1+\Omega_{k})^{5/2}}
\Big\{\Big[2(q+1)+\frac{3q(1-\Omega_{k})}{(1+\Omega_{k})}-3(1+\omega_{x}+\lambda)\Big]
\lambda(1+\Omega_{k})\Omega_{x}\nonumber\\+\Big[2q(q+1)-\frac{5q^2\Omega_{k}}{(1+\Omega_{k})}
+\frac{\dot{q}}{H}-3q\Big]\Omega_{m}\Big\},\label{eqddQm}
\end{eqnarray}
where
\begin{equation}
\dot{q}=Hq\Omega_{k}+
\frac{3}{2}H\Omega_{x}\omega_{x}\Big[2(q+1)-3(1+\omega_{x}+\lambda)\Big].\label{eqdotq}
\end{equation}
Using Eqs. (\ref{eqTx}), (\ref{eqTm}), (\ref{eqdQx}),
(\ref{eqdQm}), (\ref{eqddQx}) and (\ref{eqddQm}), we obtain
\begin{eqnarray}
\dot{Q}_{x}\ddot{Q}_{x}=\frac{9H\Omega_{x}^2 }{4(1+\Omega_{k})^5}
\Big\{\Big[2(q+1)-\frac{3q\Omega_{k}}{(1+\Omega_{k})}-3(1+\omega_{x}+\lambda)\Big]
\Big(q(1+\omega_{x})-\lambda(1+\Omega_{k})\Big)^2\nonumber\\+\Big[\frac{\dot{q}}{H}-\frac{2q^2\Omega_{k}}{(1+\Omega_{k})}\Big](1+\omega_{x})
\Big(q(1+\omega_{x})-\lambda(1+\Omega_{k})\Big)\Big\},\label{QxQx}
\end{eqnarray}
\begin{eqnarray}
\dot{Q}_{h}\ddot{Q}_{h}=\frac{9Hq}{4(1+\Omega_{k})^5}
\Big\{\Big[\frac{(2-3\Omega_{k})}{(1+\Omega_{k})}q^2+\frac{\dot{q}}{H}-q\Big]
(1+\Omega_{k}+\Omega_{x}\omega_{x})^2\nonumber\\-3q\Omega_{x}\omega_{x}(1+\omega_{x}+\lambda)
(1+\Omega_{k}+\Omega_{x}\omega_{x})\Big\},\label{QhQh}
\end{eqnarray}
\begin{eqnarray}
\frac{\dot{Q}_{m}}{T_{m}}+\frac{\dot{Q}_{x}}{T_{x}}=\frac{3}{2T_{x}(1+\Omega_{k})^{5/2}}
\Big\{\frac{r_{eq}}{r}e^{2(x-x_{eq})}\Big[\lambda(1+\Omega_{k})\Omega_{x}+q(1+\Omega_{k}-\Omega_{x})\Big]\nonumber\\
+\Big[q(1+\omega_{x})-\lambda(1+\Omega_{k})\Big]\Omega_{x}\Big\}.\label{Qh}
\end{eqnarray}

Using $x-x_{eq}=\ln{(a/a_{eq})}=\ln\Big(\frac{1+z_{eq}}{1+z}\Big)$
and Eqs. (\ref{eqTx}) and (\ref{ah}), one can rewrite Eq.
(\ref{Qh}) as
\begin{eqnarray}
\frac{\dot{Q}_{m}}{T_{m}}+\frac{\dot{Q}_{x}}{T_{x}}=\frac{3\pi
}{H(1+\Omega_{k})^3}\Big(\frac{1+z_{eq}}{1+z}\Big)^{3(\omega_{x}+\lambda)}
\Big\{\frac{r_{eq}}{r}\Big(\frac{1+z_{eq}}{1+z}\Big)^2\Big[\lambda(1+\Omega_{k})\Omega_{x}+q(1+\Omega_{k}-\Omega_{x})\Big]\nonumber\\
+\Big[q(1+\omega_{x})-\lambda(1+\Omega_{k})\Big]\Omega_{x}\Big\}.\label{Qh1}
\end{eqnarray}

Using Eqs. (\ref{q2}), (\ref{SAah}), (\ref{eqdotq}), (\ref{QxQx}),
(\ref{QhQh}), and (\ref{Qh1}), and taking $\omega_x=-1$,
$\lambda=0.3$, $z_{eq}=5.56\times10^7$, $r_{eq}=1.09\times10^5$
\cite{Zhou1}, $\Omega_{x}=0.72$, and $\Omega_{k}=0.02$ for the
present time, i.e. $z=0$, we get
$$q=-0.57,$$ $$\dot{q}=0.03H,$$
and
$$\frac{\dot{Q}_{m}}{T_{m}}+\frac{\dot{Q}_{x}}{T_{x}}=1.91\times10^4H^{-1},$$
$$\dot{Q}_{x}\ddot{Q}_{x}=-6.40\times10^{-4}H,$$
$$\dot{Q}_{h}\ddot{Q}_{h}=11.76\times10^{-4}H,$$
$$\dot{S}_{h}=2.72H^{-1}.$$

Therefore, the GSL in irreversible thermodynamics, i.e. Eq.
(\ref{eqnonStot}), for the present time yields
\begin{eqnarray}
\dot{S}^*&=&H^{-1}\Big(1.91\times10^4+6.40\times10^{-4}A_xH^2-11.76\times10^{-4}A_hH^2+2.72\Big),\nonumber\\
&=&H^{-1}\Big(1.91\times10^4+6.40\times10^{-4}\bar{A}_x-11.76\times10^{-4}\bar{A}_h+2.72\Big),\label{eqnonStot1}
\end{eqnarray}
where we define $\bar{A}:=AH^2$. If we set $\bar{A}_x=\bar{A}_h$,
then Eq. (\ref{eqnonStot1}) shows that when
$\bar{A}_x=\bar{A}_h\leq 3.56\times 10^7$, the GSL in
nonequilibrium thermodynamics is respected, i.e. $\dot{S}^*\geq
0$, for the present time.
\section{Conclusions}
Here the GSL in nonequilibrium thermodynamics for the interacting
DE with CDM in a non-flat FRW universe is investigated. The
boundary of the universe is assumed to be enveloped by the
dynamical apparent horizon. The dynamical apparent horizon in
comparison with the cosmological event horizon, is a good boundary
for studying cosmology, since on the apparent horizon there is the
well known correspondence between the first law of thermodynamics
and the Einstein equation \cite{Gong5}. In the other words, the
Friedmann equations describe local properties of spacetimes and
the apparent horizon is determined locally, while the cosmological
event horizon is determined by global properties of spacetimes
\cite{Cai05}. We assumed that when the DE and DM evolve
separately, each of them remain in thermal equilibrium with the
Hawking temperature on the dynamical apparent horizon. We found
that for the present time, the GSL in irreversible thermodynamics
is respected
 for the special range of the energy transfer
constants.
\\
\\
\noindent{{\bf Acknowledgements}}. This work has been supported
financially by Research Institute for Astronomy $\&$ Astrophysics
of Maragha (RIAAM), Maragha, Iran.



\begin{thebibliography}{}

\bibitem{Riess} A.G. Riess et al., Astron. J. {\bf 116}, 1009 (1998); S. Perlmutter et
al. Astrophys. J. {\bf 517}, 565 (1999); P. de Bernardis et al.,
Nature {\bf 404}, 955 (2000); S. Perlmutter et al., Astrophys. J.
{\bf 598}, 102 (2003).

\bibitem{Weinberg} S. Weinberg, Rev. Mod. Phys. {\bf 61}, 1
(1989).

\bibitem{Copeland} E.J. Copeland, M. Sami, S. Tsujikawa, Int. J.
Mod. Phys. D {\bf 15}, 1753 (2006).

\bibitem{Wetterich} C. Wetterich, Nucl. Phys. B {\bf 302}, 668
(1988); B. Ratra, J. Peebles, Phys. Rev. D {\bf 37}, 321 (1988).

\bibitem{Caldwell1} R.R. Caldwell, Phys. Lett. B {\bf 545}, 23
(2002); S. Nojiri, S.D. Odintsov, Phys. Lett. B {\bf 562}, 147
(2003); S. Nojiri, S.D. Odintsov, Phys. Lett. B {\bf 565}, 1
(2003).

\bibitem{Chiba} T. Chiba, T. Okabe, M. Yamaguchi, Phys. Rev. D {\bf 62},
023511 (2000); C. Armend\'{a}riz-Pic\'{o}n, V. Mukhanov, P.J.
Steinhardt, Phys. Rev. Lett. {\bf 85}, 4438 (2000); C.
Armend\'{a}riz-Pic\'{o}n, V. Mukhanov, P.J. Steinhardt, Phys. Rev.
D {\bf 63}, 103510 (2001).

\bibitem{Picon3} C. Armend\'{a}riz-Pic\'{o}n, T. Damour, V.
Mukhanov, Phys. Lett. B {\bf 458}, 209 (1999); J. Garriga, V.
Mukhanov, Phys. Lett. B {\bf 458}, 219 (1999).

\bibitem{Sen} A. Sen, J. High Energy Phys. {\bf 04}, 048 (2002); T. Padmanabhan, Phys. Rev. D {\bf 66},
021301 (2002); T. Padmanabhan, T.R. Choudhury, Phys. Rev. D {\bf
66}, 081301 (2002).

\bibitem{Gasperini} M. Gasperini, F. Piazza, G. Veneziano, Phys. Rev. D {\bf 65},
023508 (2002); N. Arkani-Hamed, P. Creminelli, S. Mukohyama, M.
Zaldarriaga, J. Cosmol. Astropart. Phys. {\bf 04}, 001 (2004); F.
Piazza, S. Tsujikawa, J. Cosmol. Astropart. Phys. {\bf 07}, 004
(2004).

\bibitem{Elizalde} E. Elizalde, S. Nojiri, S.D. Odinstov, Phys.
Rev. D {\bf 70}, 043539 (2004); S. Nojiri, S.D. Odintsov, S.
Tsujikawa, Phys. Rev. D {\bf 71}, 063004 (2005); A. Anisimov, E.
Babichev, A. Vikman, J. Cosmol. Astropart. Phys. {\bf 06}, 006
(2005).

\bibitem{Kamenshchik} A. Kamenshchik, U. Moschella, V. Pasquier, Phys. Lett. B {\bf 511}, 265
(2001); M.C. Bento, O. Bertolami, A.A. Sen, Phys. Rev. D {\bf 66},
043507 (2002).

\bibitem{Deffayet} C. Deffayet, G.R. Dvali, G. Gabadadze, Phys. Rev. D {\bf 65},
044023 (2002); V. Sahni, Y. Shtanov, J. Cosmol. Astropart. Phys.
{\bf 11}, 014 (2003).

\bibitem{Cohen} A. Cohen, D. Kaplan, A. Nelson, Phys. Rev. Lett. {\bf 82}, 4971
(1999); P. Horava, D.Minic, Phys. Rev. Lett. {\bf 85}, 1610
(2000); S.D. Thomas, Phys. Rev. Lett. {\bf 89}, 081301 (2002); M.
Li, Phys. Lett. B {\bf 603}, 1 (2004); A. Sheykhi, Phys. Lett. B
{\bf 681}, 205 (2009);  K. Karami, J. Fehri, [arXiv:0911.4932]; K.
Karami, J. Fehri, accepted for publication in Phys. Lett. B (2010)
[arXiv:0912.1541].

\bibitem{Cai} R.G. Cai, Phys. Lett. B {\bf 657}, 228 (2007); H. Wei, R.G. Cai, Phys. Lett. B
{\bf 660}, 113 (2008); K.Y. Kim, H.W. Lee, Y.S. Myung, Phys. Lett.
B {\bf 660}, 118 (2008); H. Wei, R.G. Cai, Phys. Lett. B {\bf
663}, 1 (2008); J.P. Wu, D.Z. Ma, Y. Ling, Phys. Lett. B {\bf
663}, 152 (2008); Y.W. Kim et al., Mod. Phys. Lett. A {\bf 23},
3049 (2008); J. Zhang, X. Zhang, H. Liu, Eur. Phys. J. C {\bf 54},
303 (2008); H. Wei, R.G. Cai, Eur. Phys. J. C {\bf 59}, 99 (2009);
I.P. Neupane, Phys. Lett. B {\bf 673}, 111 (2009);  A. Sheykhi,
[arXiv:0908.0606]; A. Sheykhi, Phys. Lett. B {\bf 680}, 113
(2009); K. Karami, M.S. Khaledian, F. Felegary, Z. Azarmi,
[arXiv:0912.1536].

\bibitem{Huang} Q.G. Huang, M. Li, J. Cosmol. Astropart. Phys. {\bf 08}, 013 (2004).

\bibitem{Bennett} C.L. Bennett et al., Astrophys. J. Suppl. {\bf 148}, 1 (2003); D.N.
Spergel, Astrophys. J. Suppl. {\bf 148}, 175 (2003); M. Tegmark et
al., Phys. Rev. D {\bf 69}, 103501 (2004); U. Seljak, A. Slosar,
P. McDonald, J. Cosmol. Astropart. Phys. {\bf 10}, 014 (2006);
D.N. Spergel et al., Astrophys. J. Suppl. {\bf 170}, 377 (2007).

\bibitem{Hawking} S.W. Hawking, Commun. Math. Phys. {\bf 43}, 199 (1975).

\bibitem{Bardeen} J.M. Bardeen, B. Carter and S.W. Hawking, Commun. Math. Phys. {\bf
31}, 161 (1973).

\bibitem{Jacobson} T. Jacobson, Phys. Rev. Lett. {\bf 75}, 1260 (1995).


\bibitem{Cai05} R.-G. Cai, S.P. Kim, J. High Energy Phys. {\bf 02}, 050
(2005).

\bibitem{Akbar} M. Akbar, R.-G. Cai, Phys. Rev. D {\bf 75}, 084003 (2007).

\bibitem{Davies} P.C.W. Davies, Class. Quantum Grav. {\bf 4}, L225
(1987).


\bibitem{Gong} Y. Gong, B. Wang, A. Wang, J. Cosmol. Astropart. Phys. {\bf
01}, 024 (2007).

\bibitem{Izquierdo2} G. Izquierdo, D. Pavon, Phys. Lett. B {\bf 639}, 1
(2006); H. Mohseni Sadjadi, Phys. Lett. B {\bf 645}, 108 (2007).

\bibitem{Izquierdo} G. Izquierdo, D. Pavon, Phys. Lett. B {\bf 633}, 420 (2006).

\bibitem{Zhau} J. Zhou, B. Wang, Y. Gong, E. Abdalla, Phys. Lett. B {\bf
652}, 86 (2007).

\bibitem{Gong5} Y. Gong, B. Wang, A. Wang, Phys. Rev. D {\bf 75},
123516 (2007).

\bibitem{Wang2} B. Wang, Y. Gong, E. Abdalla, Phys. Rev. D {\bf 74}, 083520
(2006).

\bibitem{Sheykhi3} A. Sheykhi, B. Wang, Phys. Lett. B {\bf 678},
434 (2009); A. Sheykhi, B. Wang, [arXiv:0811.4477].

\bibitem{Sheykhi4} A. Sheykhi, [arXiv:0910.0510].

\bibitem{Sheykhi2} A. Sheykhi, J. Cosmol. Astropart. Phys. {\bf
05}, 019 (2009).

\bibitem{KA} K. Karami, A. Abdolmaleki, [arXiv:0909.2427].

\bibitem{Karami} K. Karami, accepted for publication in J. Cosmol. Astropart. Phys. (2010)
[arXiv:0911.4808].


\bibitem{Hu} B. Hu, Y. Ling, Phys. Rev. D {\bf 73},
123510 (2006).

\bibitem{Das} S. Das, P. Majumdar, R.K. Bhaduri, Class. Quantum Grav.
{\bf 19}, 2355, (2002).

\bibitem{Wang4} B. Wang, C.Y. Lin, D. Pav\'{o}n, E. Abdalla, Phys.
Lett. B {\bf 662}, 1 (2008).

\bibitem{Pavon} D. Pav\'{o}n, B. Wang, Gen. Relativ. Gravit. {\bf 41}, 1
(2009).

\bibitem{Zhou1} J. Zhou, B. Wang, D. Pav\'{o}n, E. Abdalla, Mod.
Phys. Lett. A {\bf 24}, 1689, (2009).

\bibitem{Tsujikawa} S. Tsujikawa, M. Sami, Phys. Lett. B {\bf 603}, 113
(2004).

\bibitem{Amendola} L. Amendola, Phys. Rev. D
{\bf 60}, 043501 (1999); L. Amendola, Phys. Rev. D {\bf 62},
043511 (2000); B. Wang, Y. Gong, E. Abdalla, Phys. Lett. B {\bf
624}, 141 (2005); D. Pavon, W. Zimdahl, Phys. Lett. B {\bf 628},
206 (2005);  M. Szydlowski, Phys. Lett. B {\bf 632}, 1 (2006); H.
Kim, H.W. Lee, Y.S. Myung, Phys. Lett. B {\bf 632}, 605 (2006); S.
Tsujikawa, Phys. Rev. D {\bf 73}, 103504 (2006);  Z.K. Guo, N.
Ohta, S. Tsujikawa, Phys. Rev. D {\bf 76}, 023508 (2007); G.
Caldera-Cabral, R. Maartens, L.A. Ure\~{n}a-L\'{o}pez, Phys. Rev.
D {\bf 79}, 063518 (2009); K. Karami, S. Ghaffari, J. Fehri, Eur.
Phys. J. C {\bf 64}, 85 (2009).

\bibitem{Wang8}
B. Wang, Ch.-Y. Lin, E. Abdalla, Phys. Lett. B {\bf 637}, 357
(2006); B. Wang, J. Zang, Ch.-Y. Lin, E. Abdalla, S. Micheletti,
Nucl. Phys. B {\bf 778}, 69 (2007); C. Feng, B. Wang, Y. Gong,
R.-K. Su, J. Cosmol. Astropart. Phys. {\bf 09}, 005 (2007).

\bibitem{Bertolami8} O. Bertolami, F. Gil Pedro, M. Le Delliou, Phys.
Lett. B {\bf 654}, 165 (2007); O. Bertolami, F. Gil Pedro, M. Le
Delliou, Gen. Rel. Grav. {\bf 41}, 2839 (2009);  E. Abdalla, L.R.
Abramo, L. Sodre, B. Wang, Phys. Lett. B {\bf 673}, 107 (2009).

\bibitem{Cai09} R.-G. Cai, L.-M. Cao, Y.-P. Hu, Class. Quantum Grav. {\bf 26}, 155018
(2009).

\end{thebibliography}
\end{document}